\documentclass[pra,amsmath,amssymb]{revtex4}
\usepackage{epsfig}
\usepackage{mathrsfs}
\usepackage[hypertex,linkcolor=red]{hyperref}
\def\SU{\ensuremath{\mathbb{SU}}}
\def\SO{\ensuremath{\mathbb{SO}}}
\def\Span{\operatorname{Span}}
\def\grp#1{{\mathbf #1}}
\def\Cmplx{\ensuremath{\mathbb C}}
\def\Reals{\ensuremath{\mathbb R}}
\def\Tr{\ensuremath{\mathrm{Tr}}}
\def\d{\ensuremath{\mathrm{d}}}
\def\spc#1{\mathcal{#1}}
\def\<{\ensuremath{\langle}}
\def\>{\ensuremath{\rangle}}
\def\kk{\ensuremath{\rangle\!\rangle}}
\def\bb{\ensuremath{\langle\!\langle}}
\def\Bnd#1{\ensuremath{\mathcal{B}(#1)}} 
\def\N#1{\ensuremath{|\!|#1|\!|}} 
\newtheorem{theo}{Theorem}
\def\Proof{\medskip\par\noindent{\bf Proof. }}
\begin{document}
\title{Applications of the group $\SU(1,1)$ for quantum computation and tomography}

\author{Giulio Chiribella}
\author{Giacomo M. D'Ariano}
\author{Paolo Perinotti}
\affiliation{Dipartimento di Fisica ``A. Volta'' and CNISM, via Bassi 6,
I-27100 Pavia, Italy.}

\begin{abstract}
  This paper collects miscellaneous results about the group $\SU(1,1)$
  that are helpful in applications in quantum optics. Moreover, we
  derive two new results, the first is about the approximability of
  $\SU(1,1)$ elements by a finite set of elementary gates, and the
  second is about the regularization of group identities for
  tomographic purposes.
\end{abstract}
\maketitle
\section{introduction}

In the last decades many achievements in quantum optics came from
nonlinear effects in crystals (for a review on the topic see
\cite{boyd}).  Nonlinear crystals allowed to produce both single mode
squeezed states, which carry attenuated quadrature noise and
constitute good carriers for classical information
\cite{yuen,walls,caves}, and two-modes squeezed states, such as the
twin beam, which is a prototype for harmonic oscillator entangled
states and are useful in many applications, such as continuous
variables teleportation \cite{braunkim}. On mathematical grounds, the
action of nonlinear crystals can be described by parametric unitary
transformations in which the pump mode is considered as a classical
field and its creation and annihilation operators are substituted by
the complex amplitude. The effective Hamiltonian allows the parametric
down conversion, which is the process by which a photon with high
frequency is annihilated and two photons with lower frequencies are
created.  This process gives rise to time evolution that can be
described through unitaries in the Schwinger representation of the
group $\SU(1,1)$, namely exponentials of linear combinations of the
three generators
\begin{equation}
  K_+=a^\dag b^\dag,\quad K_-=ab,\quad K_z=\frac12(a^\dag a+b^\dag b+1),
\end{equation}
where $a$ and $b$ are the annihilation operators for the two modes.
The degenerate parametric down conversion happens when the two created
photons are in the same mode with a frequency which is half of the
annihilated photon frequency, and this particular case giving rise to
single mode squeezing corresponds to $a=b$ with the three generators
\begin{equation}
  K_+=\frac12(a^\dag)^2,\quad K_-=\frac12a^2,\quad K_z=\frac12(a^\dag a+1/2).
\end{equation}\par

Squeezed states and twin beams are nowadays widely used in
experimental quantum optics, and it is clear that the ability of
manipulating radiation modes by unitaries of the group $\SU(1,1)$ is
crucial. In this paper we consider some general aspects of the group
$\SU(1,1)$ that can be exploited on the physical ground in order to
approximately simulate any $\SU(1,1)$ transformation by a finite set
of elementary {\em gates}, namely unitary transformations which can be
applied in a given succession in order to approach a target unitary in
the representation of $\SU(1,1)$. This is very useful in a situation
in which an experimenter needs a flexible setup which allows to
simulate within some accuracy any possible gate. A similar situation
holds for qubits, where a very powerful theorem due to Solovay and
Kitaev states that any gate can be efficiently approximated by a
finite set of elementary gates. In the case of harmonic oscillators,
however, the theorem still lacks an important part, which states that
the amount of elementary gates needed in order to approximate any gate
grows logarithmically with the accuracy. This fact is due to the
dimension of the Hilbert space, and some intermediate result toward
the analog of the qubit Solovay-Kitaev theorem in the case of harmonic
oscillators can be derived with the reasonable assumption that the
states of interest on which the gates have to be applied have finite
average energy and finite variance of the energy distribution. In the
paper we will also discuss severe limitations that forbid to find a
power law which is independent of the group element that one wants to
approximate.\par

Besides the problem of approximation of squeezed states we can
consider the problem of classifying and analyzing the performances of
covariant measurements and tomographic measurements. The first ones
are an idealization of physical measurements which turns out to be
interesting because they saturate bounds on precision for the
estimation of squeezing parameters, thus providing an absolute
standard for rating of actual detectors. As regards the tomographic
measurements, their statistics allows to completely determine the
state of radiation modes---up to statistical errors. In Ref.
\cite{macca} a particular tomographic measurement has been proposed
for states with even or odd parity, based on properties of the
Schwinger representation of $\SU(1,1)$. In this paper we will discuss
the possibility of deriving similar tomographic identities from group
integrals. Moreover, an interesting mechanism because of which the
``natural'' group integral does not converge for physically interesting
representations, and a sort of regularization is needed is shown.
This analysis provides a whole range of tomographic POVMs
corresponding to different regularizations, which can be studied in
order to optimize the performances of $\SU(1,1)$ tomography. The
technique is general and can be applied to many tomographic
measurements originated from other groups. The core of the
regularization technique consists in modifying the invariant (Haar)
measure on the group manifold, and this modification gives rise to a
generalization of the {\em Duflo-Moore} \cite{duflom,carey} operator
which is typical in groups which are not unimodular, namely for which
the invariant Haar measure does not exist. This fact implies some complication
in the data processing with respect to the usual homodyne tomography,
but on the other hand allows to optimize the group measure in order to
minimize the statistical errors.\par

In Section \ref{gen} we discuss some general aspects of the group
$\SU(1,1)$, considering its defining representation. The results
derived there will be exploited in subsequent sections. In Section
\ref{elga} we prove the existence of a set of three elementary gates,
and discuss the possibility to use them for approximation of target
group elements under reasonable assumptions on the physical states. We
also discuss the impossibility of having the exact analog of the
Solovay-Kitaev theorem for the quantum optical representations of
$\SU(1,1)$. In Section \ref{tomo} we show that the physical
representations of $\SU(1,1)$ are not square summable, and we show how
one can modify the group theoretical identities for group integrals in
order to obtain converging integrals which are useful for group
tomography. In Section \ref{conc} we close the paper with a summary of
the contents and concluding remarks.

\section{General aspects of the group $\SU(1,1)$}\label{gen}

$\SU (1,1)$ is the group of complex $2 \times 2$ matrices $M$ with
unit determinant that satisfy the relation
\begin{equation}\label{definingEq}
M^\dag P M = P~, 
\end{equation}
where 
\begin{equation}
P = \begin{pmatrix} 1&0 \\ 0 &-1 
\end{pmatrix}~.   
\end{equation}  
This relation implies that the elements of $\SU(1,1)$ preserve the
Hermitian form $\omega (v_1,v_2) \doteq v_1^{\dag} P v_2$ for
arbitrary column vectors $v_i \in \Cmplx^2.$

From the above definition it is simple to show that any matrix $M \in
\SU(1,1)$ has the form
\begin{equation}\label{Parametrization1}
M = \left( 
\begin{array}{ll}
\alpha & \bar \beta\\
\beta & \bar \alpha
\end{array} 
\right) 
\end{equation}
for $\alpha, \beta$ complex numbers such that $ |\alpha|^2 -|\beta|^2
=1$.  Notice that the columns $M_1,M_2$ of $M$ are orthogonal and
normalized with respect to the form $\omega$, namely
$\omega(M_1,M_2)=0,~ \omega(M_1,M_1)=1$, and $\omega(M_2,M_2)=-1$.  By
writing $\alpha = t + i z$ and $\beta = x +i y$, we obtain
\begin{equation}\label{Parametrization2}
M = t \openone + i z \sigma_z  + x \sigma_x + y \sigma_y~,  \qquad  t^2 + z^2 -x^2 -y^2 =1~,
\end{equation} 
$\openone$ and $\sigma_x,\sigma_y,\sigma_z$ being the identity and the
three Pauli matrices, respectively.  In other words, the elements of
$\SU(1,1)$ are parametrized by points of an hyperboloid in $\Reals^4$.
This makes $\SU(1,1)$ a Lie group, namely a group which is also a
differentiable manifold. The above parametrization clearly exhibits
three relevant facts: \emph{i)} a group element is in one-to-one
correspondence with three real parameters ($x, y$ and $z$, for
example), namely the group manifold is three dimensional \emph{ii)}
the group $\SU(1,1)$ is not compact, and \emph{iii)} it is not simply
connected\cite{NotSimplyConnected}.\par

Given a parametrization $M(\vec r)$, where the element $M(\vec r) \in
\SU(1,1)$ is specified by the triple $\vec r \in \Reals^3$, the matrix
multiplication induces a composition law in the parameter space:
$(\vec r, \vec s) \mapsto \vec r \circ \vec s$, where $\vec r \circ
\vec s$ is defined by the relation: $M(\vec r \circ \vec s)= M(\vec r)
M(\vec s)$. In particular, if $\vec r= (x,y,z)$, with $x,y,z$ as in
Eq. (\ref{Parametrization2}), we can define the invariant measure
\begin{equation}\label{InvMeas}
\d \mu (\vec r) =\frac{1}{\sqrt{1+x^2+y^2-z^2}} \d x \d y \d z~.
\end{equation}  
Invariance of the measure means that the action of the group does not
change the volume of regions in the parameter space, namely, for any
$\vec r, \vec s$, $\d \mu (\vec r \circ \vec s)= \d \mu (\vec s \circ
\vec r) =\d \mu (\vec r)$.  The expression (\ref{InvMeas}) of the
invariant measure $\d \mu(x,y,z)$ is particularly useful, since it
allows to obtain the invariant measure in any parametrization of the
group, just by performing a change of variables. For example, a useful
alternative parametrization of a group element $M \in \SU(1,1)$ is
given by
\begin{equation}\label{Parametrization3}
M(\theta, \phi, \psi)= \left(
\begin{array}{ll}
  \cosh \theta~e^{i \phi} & \sinh \theta~ e^{-i \psi}\\
  \sinh \theta~e^{i \psi} & \cosh \theta~ e^{-i \phi}
\end{array}
\right)~,
\end{equation}
for $\theta \in [0, +\infty), \phi \in [0,2 \pi), \psi \in [0,2\pi)$.
The change of parametrization from (\ref{Parametrization2}) to
(\ref{Parametrization3}) corresponds to the change of variables $x=
\sinh \theta \cos \psi~, y= \sinh \theta \sin \psi~, z = \cosh \theta
\sin \phi$. Performing the change of variables in Eq. (\ref{InvMeas})
we obtain the expression of the invariant measure in the
parametrization $M=M(\theta, \phi, \psi)$, namely
\begin{equation}\label{InvMeas2}
  \d \nu (\theta, \phi, \psi) = \sinh \theta \cosh \theta ~ \d \theta \d \phi \d \psi~. 
\end{equation}\par

\subsection{The Lie algebra $su(1,1)$}
Since $\SU(1,1)$ is a real three-dimensional manifold, its Lie algebra
$su(1,1)$---the tangent space in the identity---is a three-dimensional
vector space. As usual, a basis of the Lie algebra is obtained by
differentiating curves passing through the identity.  Differentiation
with respect to the parameters $x,y,z$ in the identity provides the
generators
\begin{align}
  i \sigma_x &= i \left[ \frac{\d}{\d x} M(x,y,z)
  \right]_{x=y=z=0},\\
  i \sigma_y &= i \left[ \frac{\d}{\d y} M(x,y,z)
  \right]_{x=y=z=0},\\
  -\sigma_z &= i \left[ \frac{\d}{\d z} M(x,y,z) \right]_{x=y=z=0},
\end{align}
where $M(x,y,z)$ is defined by Eq.~\eqref{Parametrization2}. Hence the
Lie algebra $su(1,1)$ is the real vector space spanned by the matrices
$i\sigma_x, i\sigma_y$, and $\sigma_z$. By defining $k_{x}= i
\frac{\sigma_{x}}{2}$, $k_y = i \frac{\sigma_y}{2}$, $k_z=
\frac{\sigma_z}{2}$, and $k_{\pm} = k_x \pm i k_y$, we obtain the
standard commutation relations
\begin{equation}\label{CommRel}
\left\{ \begin{array}{lll} 
{[k_{+},k_{-}]} &=& {-2k_z}\\
&&\\
{[k_z,k_{\pm}]} &=& {\pm k_{\pm}}~.
\end{array}
\right.
\end{equation}
By definition, an operator representation of the algebra $su (1,1)$ is
given by the assignment of three operators $K_x, K_y$ and $K_z$ that
satisfy the above commutation relations with $K_{\pm}= K_x \pm i K_y$.
From such relations, it follows that in any representation of
$su(1,1)$ the Casimir operator
\begin{equation}\label{Casimir}
\vec K \cdot \vec K \doteq K^2_z-K_x^2 -K_y^2
\end{equation}
commutes with  the whole algebra spanned by $K_x, K_y, K_z$.\par

\subsection{The exponential map}\label{ExpoMap}
A way of writing the group elements in any representation in terms of
the Lie algebra generators is through the exponential map. The
exponential map $M= e^{i m}$ is the map that associates an element $m
\in su(1,1)$ of the Lie algebra with an element $M \in \SU(1,1)$ of
the group. In order to discuss the exponential map, it is suitable to
write the elements of the algebra as $m= \chi~ \vec n \cdot \vec k$,
where $\chi \in \Reals$, $\vec n \cdot \vec k \doteq n_z k_z-n_x k_x
-n_y k_y$ and $\vec n\in \Reals^3$ is a normalized vector. In this
context \emph{normalized} means that the product $\vec n \cdot \vec n
\doteq n_z^2- n_x^2- n_y^2$ can assume only the values $+1$, $-1$, and
$0$.  Then, the exponentiation of the element $m \in su(1,1)$ is
easily performed by using the relation
\begin{equation}\label{Square}
  \left( \vec n \cdot \vec k \right)^2 = \frac{\vec n \cdot \vec n}{4}~\openone~, 
\end{equation}
which follows directly from the properties of Pauli matrices.  In the
following, we analyze the three cases $\vec n\cdot \vec n=\pm 1,0$
separately.
 
{\bf Case 1: $\vec n \cdot \vec n= +1$.}  The exponentiation gives
\begin{equation}\label{Element+}
  M_+= e^{i \chi \vec n\cdot \vec k} = \cos \left (\frac \chi 2\right)~
  \openone + i \sin \left( \frac \chi 2 \right)~ 2\vec n \cdot \vec k~.
\end{equation} 
Notice that, for any fixed direction $\vec n$, we have a one-parameter
subgroup, which is compact and isomorphic to $U(1)$.

The group elements of the form (\ref{Element+}) form a region
$\Omega_+ \subset \SU(1,1)$, which contains $\pm \openone$ and all
the matrices $M \in \SU(1,1)$ such that $\left|\Tr [M]\right| <2$.

{\bf Case 2: $\vec n\cdot \vec n= -1$.} Exponentiating the generator
$\vec n\cdot \vec k$ we obtain:
\begin{equation}\label{Element-}
  M_-= e^{i \chi \vec n\cdot \vec k}= \cosh \left(\frac \chi 2 \right)~ \openone + i \sinh \left( \frac \chi 2 \right)~ 2\vec n\cdot \vec k~. 
\end{equation}
In this case, for a fixed direction $\vec n$ we have a one-parameter
subgroup, which is not compact and is isomorphic to $\Reals$.  The
elements $M_-$ form a region $\Omega_- \subset \SU(1,1)$, which
contains the identity and all the matrices $M\in \SU(1,1)$ such that
$\Tr [M] >2$.

{\bf Case 3: $\vec n \cdot \vec n=0$.} In this case, the
exponentiation gives
\begin{equation}
  M_0 = e^{i \chi \vec n \cdot \vec k} = \openone + i \chi \vec n \cdot
  \vec k~.
\end{equation}
The elements $M_0$ form a region $\Omega_0 \subset \SU(1,1)$, which
contains all matrices $M \in \SU(1,1)$ such that $\Tr [M]=2$.  The
region $\Omega_0$ is a two-dimensional surface, and therefore, it has
zero volume.\par

We want to stress that the exponential map does not cover the whole
group $\SU(1,1)$. The region of $\SU(1,1)$ covered by the exponential
map is $\Omega = \Omega_+ ~\cup~ \Omega_- ~\cup ~\Omega_0$, and
contains matrices $M \in \SU(1,1)$ such that $\Tr[M] \ge -2$.
However, according to the parametrization (\ref{Parametrization2}),
the trace of a matrix $M \in \SU(1,1)$ is $\Tr [M]= 2 t$, $t \in
\Reals$. Therefore, the group $\SU(1,1)$ contains also elements with
trace $\Tr [M] < -2$, that cannot be obtained with the exponential map.
Nevertheless, any matrix $M \in \SU(1,1)$ with $\Tr [M] < -2$ can be
written as $M = - M_-$, for some $M_- \in \Omega_-$, and any matrix $M
\in \SU(1,1)$ with $\Tr [M]=-2$ can be written as $M= -M_0$ for some
$M_0 \in \Omega_0$. Defining $-\Omega_- \doteq \{-M_-~|~ M_- \in \Omega_-\}$ and $-\Omega_0 \doteq \{-M_0~|~M_0 \in \Omega_0\}$ we have
\begin{equation}
\SU (1,1) = \Omega \cup -\Omega_- \cup -\Omega_0~.
\end{equation}
 Notice that, since $\Omega_0$ and $-\Omega_0$ have zero measure, any group integral can be written as the sum of only three contributions, coming from $\Omega_+, \Omega_-$, and $-\Omega_-$, respectively.

Even though the exponential map does not
cover the whole group $\SU(1,1)$, any group element $M(\theta,\phi,\psi)$---parametrized as in Eq. (\ref{Parametrization3})---can be written  as a \emph{product} of exponentials, for example as
\begin{equation}\label{Decomposition2}
  M(\theta,\phi,\psi)=
  e^{\xi k_+ -\bar \xi k_-}~ e^{2 i\phi k_z} \qquad \qquad \xi =-i
  \theta e^{-i(\psi-\phi)}~.
\end{equation}

The relation (\ref{Decomposition2}) is particularly useful, since it allows to construct from any representation of the Lie algebra $su (1,1)$ a representation of the group $\SU(1,1)$.
In particular, for the physical realizations of the group $\SU (1,1)$, where the generators $k_x,k_y, k_z$ are represented by Hermitian operators $K_x, K_y, K_z$ in an infinite dimensional Hilbert space, relation (\ref{Decomposition2}) provides the \emph{unitary} representation
\begin{equation}\label{UnitaryRep2}
U_{\theta, \phi, \psi}= e^{\xi K_+ -\bar \xi K_-}~e^{2i \phi K_z} \qquad \qquad \xi = -i \theta e^{-i (\psi-\phi)}~.
\end{equation}

\subsection{Baker-Campbell-Hausdorff formula}

The exponential with $k_+,k_-$ in Eq.~\eqref{UnitaryRep2} can be
further decomposed according to the \emph{Backer-Campbell-Hausdorff
  (BCH) formula}. The BCH formula is the fundamental relation, holding
\emph{for any} representation of the algebra $su (1,1)$, given by
\cite{puri}
\begin{equation}\label{BCH}
  e^{\xi K_+ -\bar \xi K_-}= \exp^{ \frac{\xi}{|\xi|} \tanh |\xi|~K_+}~ \left( \frac 1 {\cosh |\xi|}\right)^{2 K_z}~ \exp^{ -\frac{\bar \xi}{|\xi|} \tanh |\xi|~K_-} \qquad \forall \xi \in \Cmplx~.     
\end{equation}
This formula can be simply proved by verifying it in the case of the
two-by-two matrices $k_+, k_-, k_z \in su (1,1)$.

A version of the BCH formula in ``antinormal order'' is given by the
relation
\begin{equation}\label{BCH'}
  e^{\xi K_+ -\bar \xi K_-}= \exp^{ -\frac{\bar \xi}{|\xi|} \tanh |\xi|~K_-}~ (~\cosh |\xi|~)^{2 K_z}~ \exp^{ \frac{\xi}{|\xi|} \tanh |\xi|~K_+} \qquad \forall \xi \in \Cmplx~,
\end{equation}
which follows from (\ref{BCH}) with the change of representation
$K_+'=-K_-,~K_-'=-K_+,~ K_z'=-K_z$.

\section{elementary gates}\label{elga}

The parametrization (\ref{Parametrization3}) makes evident that any
element of $\SU(1,1)$ can be obtained as a product of exponentials of the generators
$k_z$ and $k_x$. In fact, Eq.
(\ref{Parametrization3}) is equivalent to the decomposition
\begin{equation}
  M(\theta, \phi, \psi) = e^{i (\phi-\psi) k_z}~ e^{-2i k_x}~ e^{i (\phi+ \psi) k_z} \label{Decomposition}
\end{equation} 
As a consequence, we have the following approximation theorem:
\begin{theo}[Approximation of group elements] Any element of $M \in
  \SU(1,1)$ can be approximated with arbitrary precision with a finite
  product involving only three elements $G_1, G_2, G_3 \in \SU(1,1)$.
  A possible choice is
\begin{equation}
G_1 = e^{\theta_1
\sigma_x},~ G_2= e^{-\theta_2 \sigma_x},~ G_3 = e^{i\phi_3
\sigma_z}~,
\end{equation}
with $\theta_1, \theta_2 > 0$, $\theta_1/\theta_2 \not \in \mathbb Q$,
and $\phi_3 /2 \pi \not \in \mathbb Q$.
\end{theo}
\Proof Due to decomposition (\ref{Decomposition}), it is enough to
show that all elements of the form $e^{i \phi \sigma_z}$ and of the
form $e^{\theta \sigma_x}$ can be approximated with a product of $G_1,
G_2, G_3$.  First, any point of the circle $\mathcal C= \Reals \mod 2
\pi$ can be approximated by a multiple of an angle $\phi_3$, provided
that $\phi_3$ is not rational with $2\pi$. Approximating $\phi$ as
$\phi \approx N_3 \phi_3, N_3 \in \mathbb N$ corresponds to
approximating the exponential $e^{i\phi \sigma_z}$ as $G_3^{N_3}$. In
the same way, any point of the circle $\mathcal C'=\Reals \mod
\theta_1$ can be approximated by a multiple of $-\theta_2$, provided
that $\theta_2$ is not rational with $\theta_1$. Since any real number
$\theta \in \Reals$ can be written as $\theta =M \theta_1 + \theta
\mod \theta_1$, by approximating $\theta \mod \theta_1 \approx N_2
\theta_2 \mod \theta_1$, we obtain $\theta \approx N_1 \theta_1 -N_2
\theta_2$, for some $N_1 \in \mathbb N$. This corresponds to
approximating the exponential $e^{\theta \sigma_x}$ as $G_1^{N_1}
G_2^{N_2}$.\par

The previous theorem is particularly important in consideration of
physical realizations, where the group $\SU(1,1)$ acts unitarily on an
infinite dimensional Hilbert space. In this case, the previous result
shows that any unitary transformation representing an element of
$\SU(1,1)$ can be arbitrarily approximated by a finite circuit made
only of three elementary gates. However, if we thoroughly define a
parameter for the rating of the approximation we find that the
accuracy is arbitrarily small, and this fact is due to unboundedness
of the generators for the unitary representations of $\SU(1,1)$ of
physical interest. In particular, we are interested in the two
representations in which
\begin{equation}
\begin{split}
  &K_z=\frac12(a^\dag a+b^\dag b+1),\quad K_+=a^\dag b^\dag,\quad K_-=ab,\\
  &K_z=\frac12(a^\dag a+1/2),\quad K_+=\frac12(a^\dag)^2,\quad
  K_-=\frac12a^2.
\end{split}
\end{equation}
The parameter for the approximation rating is the accuracy
$\epsilon^{-1}$, with
\begin{equation}\label{approx}
  \epsilon\doteq  \sup_{\N{\psi}=1}  \N{(U_1-U_2)|\psi\>},
\end{equation}
where $U_1$ is the target element, $U_2$ is the product of elementary
gates that approximates $U_1$. However, since we are considering
infinite dimensional representations, the difference $U_1-U_2$ has
eigenvalues arbitrarily near 2. The supremum is then always 2, and in
order to find some approximation criterion one has to impose some
constraint on the states that we are considering. For example, we will
impose that the average and second moment of the photon number
distribution are finite, which are reasonable physical assumptions.
Suppose now that we have a sufficiently long sequence of elementary
gates, in such a way that, using the decomposition of
Eq.~\eqref{Decomposition}, $U_1=e^{-i\alpha K_z}e^{-i\beta
  K_x}e^{-i\gamma K_z}$ and $U_2=e^{-i(\alpha+\delta_\alpha)K_z}
e^{-i(\beta+\delta_\beta)K_x} e^{-i(\gamma+\delta_\gamma)K_z}$, and
only first order terms in $\delta_x$ are relevant, thanks to the
constraint on states. After some algebra and exploiting
Eq.~\eqref{conjug} one can verify that the supremum of
$\<\psi|2I-U_1^\dag U_2-U_2^\dag U_1|\psi\>$ is almost equal to the
supremum of $\<\psi|\Delta|\psi\>$, where
\begin{equation}\label{delta}
  \Delta=\left\{
\begin{split}
  &(\delta_\alpha^2+\delta_\gamma^2+2\cosh\beta\delta_\alpha\delta_\gamma-\delta_\beta^2)K_z^2\\
  &(\delta_\beta^2-\delta_\alpha^2-\delta_\gamma^2-2\cosh\beta\delta_\alpha\delta_\gamma)K_x^2,
\end{split}\right.
\end{equation}
depending on the sign of
$\delta_\alpha^2+\delta_\gamma^2+2\cosh\beta\delta_\alpha\delta_\gamma
-\delta_\beta^2$. This equation implies two facts.  First of all, we
can easily verify that the physical constraint on states is necessary
in order to guarantee boundedness of $\epsilon$. However, it is not
sufficient because of the presence of $\cosh\beta$ in the expression.
This is due to non compactness of the group, which implies that even
in the defining representation the approximation is worse as one goes
further along the direction of a non compact parameter. This fact
fatally flaws any analogy to the Solovay-Kitaev theorem for the qubit
case, and in order to have a similar result one must also restrict the
set of unitaries that he wants to approximate. Otherwise, a power law
for the number of gates as a function of $\epsilon^{-1}$ can be
searched which contains an explicit dependence also on the parameter
$\beta$.  Suppose that we have
$|\delta_\alpha^2+\delta_\gamma^2+2\cosh\beta\delta_\alpha\delta_\gamma
-\delta_\beta^2|\doteq f(N)$, where $N$ is the number of elementary gates
needed to approximate the target group element with in the defining
representation.  Then, for $\Delta\propto K_z^2$ in Eq.~\eqref{delta},
we have $\epsilon=f(N)(\<E^2\>+2\lambda\<E\>+\lambda^2)$, where $E$ is
the total number of photons and $\lambda=\frac14$ for single mode
representation and $\lambda=\frac12$ for two modes. The function $f$
is clearly non increasing, and supposing that it is strictly
monotonic, it can be inverted, obtaining
\begin{equation}
  N=f^{-1}\left(\frac{\epsilon}{\<E^2\>+2\lambda\<E\>+\lambda^2}\right).
\end{equation}

\section{unitary representations of \SU (1,1)}\label{uirs}
Given a representation of the $su (1,1)$ algebra where the generators
$K_x,K_y,K_z$ are Hermitian operators acting in an infinite
dimensional Hilbert space $\spc H$, we consider the unitary representation
$U_{\theta,\phi,\psi}$ of the group $\SU (1,1)$ defined by
Eq. (\ref{UnitaryRep2}). In general, such a representation is reducible, and it can be decomposed into unitary irreducible representations (UIRs).

A UIR $U_{\theta,\phi,\psi}$ is called
\emph{square-summable} if there is a nonzero vector $|v\>\in \spc H$
such that
\begin{equation}\label{DefSquareSummable}
  \int_{\SU (1,1)} \d \nu (\theta, \phi,\psi)~ \left | \<v|~U_{\theta,\phi,\psi}~|v\> \right |^2 < \infty~,
\end{equation} 
where $\d \nu$ is the invariant measure defined in Eq.
(\ref{InvMeas2}).  Moreover, since the group $\SU(1,1)$ is unimodular,
if the above integral converges for one vector $|v\> \not =0$, then it
converges for any vector in $\spc H$ \cite{grossmorl}.\par

Square-summable representations enjoy the important property expressed
by the following:
 \begin{theo}[Formula for the group average]\label{Theo:GroupAve}
   If the irreducible representation $U_{\theta,\phi,\psi}$ is square-summable,
   then for any operator $A \in \Bnd{\spc H}$ the following relation
   holds
\begin{equation}\label{GroupAve}
\int_{\SU(1,1)}~ \d \nu (\theta,\phi,\psi)~ U_{\theta,\phi,\psi}~ A~ U_{\theta,\phi,\psi}^{\dag} = \Tr [A]~ \frac {\openone} d~.
\end{equation}
Here $\openone$ is the identity in $\spc H$, and $d$ is the
\emph{formal dimension}, defined by
\begin{equation}\label{DefFormalDimension}
d\doteq  \left( \int_{\SU(1,1)}~ \d \nu(\theta,\phi,\psi)~ |\<v|~U_{\theta,\phi, \psi}~|v\>|^2 \right)^{-1}~,  
\end{equation}
where $|v\>$ is any normalized vector $|v\> \in \spc H$, $\<v|v\>=1$.
\end{theo}
The formula for the group average is fundamental in the contexts of
quantum estimation and tomography, since it allows to construct
resolutions of the identity via a group integral. In the context of
quantum estimation, Eq. (\ref{GroupAve}) ensures that the operators
\begin{equation}\label{CovPOVM}
P(\theta,\phi,\psi)= U_{\theta,\phi,\psi}~ \xi ~U_{\theta,\phi,\psi}^\dag~,
\end{equation}
where $\xi$ is any operator satisfying $\xi \ge 0,~\Tr[\xi]=d$,
provide a \emph{positive operator valued measure (POVM)} for the joint
estimation of the three parameters $\theta,\phi,\psi$. In fact, such
operators satisfy the normalization condition
\begin{equation}
\int_{\SU(1,1)}~ \d \nu (\theta,\phi,\psi)~ P(\theta, \phi,\psi)= \openone~,
\end{equation} 
which guarantees that the total probability of all possible outcomes
is one.  In particular, if $\xi =d~ |v\>\<v|$ for some state $|v\>$,
the above formula gives the completeness of the set of $\SU(1,1)$
coherent states \cite{pere}
\begin{equation}\label{SU(1,1)CoherentState}
|v_{\theta, \phi, \psi}\> \doteq U_{\theta, \phi,\psi}~ |v\>~.
\end{equation}
\subsection{Examples}
\subsubsection{Single mode squeezing}\label{Single}
The representation of the $su (1,1)$ algebra, given by
\begin{equation}
K_+ = \frac {a^{\dag 2}}  2 \qquad K_-= \frac {a^2} 2 \qquad K_z= \frac 1 2 \left ( a^\dag a + \frac 1 2 \right) 
\end{equation}  
is reducible in the Hilbert space $\spc H$ of a single harmonic oscillator. In fact, the subspaces $\spc H_{\rm even}
=\Span \{|2n\>~|~ n \in \mathbb N\}$ and $\spc H_{\rm odd}= \Span \{
|2n +1\>~|~ n \in \mathbb N\}$, defined in terms of the Fock basis
$|n\> =\frac 1 {\sqrt{n!}} a^{\dag n}|0\>$, are invariant under the
application of $K_x, K_y, K_z$. 
The unitary representation of $\SU (1,1)$ defined by Eq. (\ref{UnitaryRep2}) acts irreducibly in the subspaces $\spc H_{\rm even}$ and
$\spc H_{\rm odd}$. There is a substantial difference between the two UIRs acting in $\spc H_{\rm odd}$ and $\spc H_{\rm even}$, in fact, the first is square-summable, with formal dimension   
 $d_{\rm odd} =1/(4 \pi^2)$, while the latter is not. A surprising consequence of the non square-summability in $\spc H_{\rm even}$ is that the squeezed states $|\xi\> = e^{\xi K_+ -\bar \xi K_-}|0\>$---which are the coherent states of $\SU (1,1)$ commonly considered in quantum optics---do not provide a resolution of the identity.

\subsubsection{Two modes squeezing}\label{Two} The representation of the Lie algebra $su (1,1)$ given by the operators
\begin{equation}
K_+ = a^\dag b^\dag, \qquad K_-=ab \qquad K_z=\frac 1 2 \left( a^{\dag} a +b^\dag b +1 \right)
\end{equation}  
is reducible in the Hilbert space $\spc H_a \otimes \spc H_b$ of two
harmonic oscillators. It is indeed immediate to see that, for any
$\delta \in \mathbb Z$, the subspaces $\spc H_{\delta} = \Span \{ |m\>
|n\>~|~m,n \in \mathbb{N}, m-n=\delta~\}$ are invariant under
application of the operators $K_+,K_-,K_z$. The unitary representation of $\SU
(1,1)$ given by Eq. (\ref{UnitaryRep2}) is irreducible in each subspace
$\spc H_{\delta}$.  The two UIRs acting in $\spc H_{\delta}$ and $\spc
H_{-\delta}$ are unitarily equivalent, while for different values of
$|\delta|$, one has inequivalent UIRs. All UIRs in the two modes
realization are square-summable, with the only exception of the case
$\delta =0$.  

%Again, it is remarkable to notice that the states $|\xi
%\>\!\> = e^{\xi K_+-\bar \xi K_-} |0\>|0\>$, which are ordinarily
%produced via parametric down-conversion in the laboratories of quantum
%optics, are coherent states of $\SU {1,1}$ for the representation
%corresponding to $\delta =0$, precisely the only representation in the
%two-mode setup that is not square-summable. Accordingly, there is no
%resolution of the identity associated to such coherent states.

\section{$\SU(1,1)$ tomography}\label{tomo}

\subsection{Reconstruction formula for square-summable representations}
Let us now consider the group as a tool for quantum tomography. In
order to do that, it is useful to consider the set of operators on a
Hilbert space $\spc H$ as a Hilbert space itself, isomorphic to $\spc
H\otimes\spc H$, and to look for spanning sets in this Hilbert space.
An immediate and handy way of defining the isomorphism between operators and bipartite vectors is through the definition
\begin{equation}
|A\kk\doteq\sum_{m,n}\<m|A|n\>~|m\>\otimes|n\>\,
\end{equation}
where $A$ is an operator on $\spc H$, and $|n\>$ are elements of a fixed basis for $\spc H$. This definition implies the
following useful identities
\begin{eqnarray}
\label{SonoFuoriDalKet}
A\otimes B|C\kk&=&|ACB^\tau\kk \\
\label{TrKet} \bb A|B\kk&=&\Tr[A^\dag B]~,
\end{eqnarray}
where $X^\tau$ denotes the transpose of $X$ in the basis $|n\>$.

For a square-summable UIR $U_{\theta, \phi,\psi}$, we can obtain a resolution of the identity by simply exploiting Eqs.~\eqref{GroupAve} and \eqref{SonoFuoriDalKet}, namely
\begin{equation}\label{GroupAveent}
\openone \otimes \openone=   d\int_{\SU(1,1)}~ \d \nu (\theta,\phi,\psi)~ |U_{\theta,\phi,\psi}\kk\bb U_{\theta,\phi,\psi}|~,
\end{equation}
which shows that the unitaries $U_{\theta, \phi,\psi}$ form a spanning set for the space of operators.

Tomographing the state $\rho$ is equivalent to reconstructing the
ensemble average $\Tr[\rho A]=\bb\rho|A\kk$ of any operator $A$ on the
state $\rho$. This can be done using the reconstruction formula
\begin{equation}\label{TomoInt}
\Tr[ \rho A] =  d\int_{\SU(1,1)}~ \d \nu (\theta,\phi,\psi)~ \Tr[\rho U_{\theta,\phi,\psi}]~ \Tr[ U_{\theta,\phi,\psi}^\dag A]~,
\end{equation}
which directly follows by inserting  the resolution of the identity (\ref{GroupAveent}) into the product $\Tr[\rho A]= \bb\rho|A\kk$.
In a real tomographic scheme, the traces $\Tr[\rho
U_{\theta,\phi,\psi}]$ have to be evaluated by experimental data,  and subsequently averaged with the {\em processing function}
$f_A(\theta,\phi,\psi)=\Tr[U^\dag_{\theta,\phi,\psi}A]$ in order to obtain the expectation value $\Tr[\rho A]$. A feasible scheme for evaluating the traces $\Tr[\rho U_{\theta, \phi,\psi}]$ by experimental data is discussed in Subsection \ref{HowTo}. \par

\subsection{Non square-summable representations: regularization}
The representations of $\SU (1,1)$ that are common in quantum optics
are single mode and two-modes Schwinger representations, analyzed in
Paragraphs. \ref{Single} and \ref{Two}, respectively.  In particular, the
irreducible subspaces $\spc H_{\rm even}$ in the single-mode case,
and $\spc H_0$ in the two-modes case are particularly interesting,
since coherent states with even photon number in the single-mode case (or, alternatively, zero difference of photon numbers in the two-modes case) are experimentally achievable by
simple vacuum squeezing.

The problem now is that, the single-mode representation in $\spc
H_{\rm even}$, and the two-modes representation in $\spc H_0$ are not
square-summable, therefore the group integral in Eq.~\eqref{DefSquareSummable}
diverges. In order to circumvent this problem, we
address here the a technique that consists in modifying the invariant
measure $\d \nu(\theta, \phi, \psi)$ by a regularization factor
$g(\theta,\phi,\psi)$, which is positive almost everywhere.  The
modification of the measure makes it non invariant, and consequently
the group average identities become similar to those of non unimodular
groups, where there is no invariant measure.

Using the regularization factor, instead of the resolution of the identity \eqref{GroupAveent},  we have a positive invertible
operator
\begin{equation}\label{frame}
  F=\int_{\SU(1,1)}~ \d \nu (\theta,\phi,\psi)~ g(\theta,\phi,\psi)~|U_{\theta,\phi,\psi}\kk\bb U_{\theta,\phi,\psi}|~,
\end{equation}
and the ensemble average of any operator $A$ can be obtained by writing $\Tr[\rho A]= \bb\rho  |F F^{-1} |A\kk$.
In this way, we can provide a \emph{regularized reconstruction formula}
\begin{equation}\label{tomid}
  \Tr[\rho A]=\int_{\SU(1,1)}\d\nu(\theta,\phi,\psi)g(\theta,\phi,\psi)f_A(\theta,\phi,\psi)~ \Tr[U_{\theta,\phi,\psi}\rho],
\end{equation}
involving the  processing function 
\begin{equation}\label{procefun}
f_A(\theta,\phi,\psi)=\bb U_{\theta,\phi,\psi}|F^{-1}|A\kk,
\end{equation}
instead of $\bb U_{\theta,\phi,\psi}|A\kk$. Notice that the identity
Eq.~\eqref{tomid} can be used also in the square-summable case with
$g(\theta,\phi,\psi)\equiv1$.

\subsection{A relevant example}

Here we consider in detail the case of the UIRs with even photon
number (single-mode case), and with $\delta=0$ (two-modes case)
representations, providing an example of the general method discussed
above. We will start from the following integral
\begin{equation}
  S(m,n;m',n')=\int_{\SU(1,1)}~ \d \nu (\theta,\phi,\psi)\<2m|U_{\theta,\phi,\psi}|2n\>\<2n'|U_{\theta,\phi,\psi}^\dag|2m'\>,
\end{equation}
where $|2m\>$ denotes both the even eigenstate of $a^\dag a$ or the
zero-difference eigenstate $|m\>|m\>$ of $a^\dag a+b^\dag b$. This can be
evaluated by exploiting Eqs.~\eqref{UnitaryRep2} and \eqref{BCH}, thus
obtaining the following expressions for the matrix element
$\<2m|U_{\theta,\phi,\psi}|2n\>$
\begin{equation}
  \<2m|U_{\theta,\phi,\psi}|2n\>=
  e^{i\phi(2n+\kappa) }e^{i(\psi-\phi)(n-m)}\sum_{p=0}^n c_\kappa(p)(-i\tanh\theta)^{2p+m-n}\left(\frac1{\cosh\theta}\right)^{2n-2p+\kappa},
\end{equation}
where $\kappa=1/2$ for even single mode states and $\kappa=1$ for
$d=0$ two modes states, and
\begin{equation}
c_\kappa(p)=\left\{
\begin{split}
&\frac{\sqrt{2n!2m!}}{p!(p+m-n)!(2n-2p)!2^{2p+m-n}}&\kappa=\frac12\\
&\frac{n!m!}{p!(p+m-n)!(n-p)!^2}&\kappa=1.
\end{split}\right.
\end{equation}
By exploiting the integral in $\psi$ and then in $\phi$ we obtain
\begin{equation}
S(m,n;m',n')=4\pi^2\delta_{m,m'}\delta_{n,n'}\sum_{p,p'=0}^nc_\kappa(p)c_\kappa(p')(-1)^{p+p'}I_\kappa(m,n,p,p'),
\end{equation}
where
\begin{equation}\label{inte}
  I_\kappa(m,n,p,p')=\int_0^\infty\d\theta\sinh\theta\cosh\theta
  \frac{(\tanh\theta)^{2(p+p')+2m-2n}}{(\cosh\theta)^{4n-2p-2p'+2\kappa}},
\end{equation}
%=\int_1^\infty\d t\frac{(1-1/t^2)^{p+p'+m-n}}{(t^2)^{2n-p-p'}}
and for $2n=p+p'$ this is clearly divergent. Moreover, for $m<n$ and
$p=p'=0$ the integral diverges because of the singularity in
$\theta=0$. If we introduce the regularization factor
$g(\theta,\phi,\psi)=g(\theta)\doteq
\frac{e^{-1/(\tanh\theta)^2}}{(\cosh\theta)^3}$, by exploiting the
same calculations we get the same result with $I_\kappa(m,n,p,p')$
substituted by
\begin{equation}
  I_{\kappa, g}(m,n,p,p')=\int_0^1\d x x^{2n-p-p'+\kappa}(1-x)^{p+p'+m-n}e^{-\frac1{1-x}},
\end{equation}
which is derived from Eq.~\eqref{inte} by the change of variable
$(1/\cosh\theta)^2\to x$, and which is finite. As a consequence
\begin{equation}\label{reggroupav}
\begin{split}
  &\int_{\SU(1,1)}\d\nu(\theta,\phi,\psi)g(\theta)|U_{\theta,\phi,\psi}\kk\bb
  U_{\theta,\phi,\psi}|=\sum_{m,n=0}^\infty
  F^{(\kappa)}_{m,n}|2m\>\<2m|\otimes|2n\>\<2n|,\\
  &F^{(\kappa)}_{m,n}=4\pi^2\sum_{p,p'=0}^n(-1)^{p+p'}c_\kappa(p)c_\kappa(p')I_{\kappa,g}(m,n,p,p')\\
  &=\int_0^1\d x\left|\sum_p(-1)^p c_\kappa(p)\left(\frac{1-x}x\right)^p\right|^2x^{2n}(1-x)^{m-n}e^{-\frac1{1-x}},
\end{split}
\end{equation}
and since the coefficients $c_\kappa(p)$ are non null, clearly
$0<F^{(\kappa)}_{m,n}<\infty$. This implies that the operator $F$ in
Eq.~\eqref{frame} is actually invertible, and we can safely use the
processing function in Eq.~\eqref{procefun} for tomographic
reconstruction of the operator $A$. Notice that this formula is very
close to the group-average formula for non unimodular groups, where
the Duflo-Moore operator $C$ is involved. The group average identity
in that case is similar to Eq.~\eqref{reggroupav} with
$F^{(\kappa)}_{m,n}=(C^\dag C)_{m,m}$.\par

\subsection{How to make tomography experimentally}\label{HowTo}
In quantum tomography, the expectation value $\<A\>= \Tr [\rho A]$ of
any observable is reconstructed by exploiting integral
\eqref{tomid}.
%\begin{equation}
%  \<A\>\simeq\frac 1N\sum_{j=1}^N ~f_A(\theta_j,\phi_j,\psi_j)\Tr[\rho U_{\theta_j,\phi_j,\psi_j}],
%\end{equation}
%where the parameters $\theta_j, \phi_j, \psi_j$ are randomly extracted according to the probability distribution
%\begin{equation}
%\d p (\theta, \phi, \psi)= \frac 1{\mathcal{N}}~ \sinh \theta \cosh \theta ~g(\theta)~ \d \theta \d \phi \d \phi~,  
%\end{equation} 
%$\mathcal{N}$ being a normalization constant.
Moreover, in order to have a feasible tomography, it is essential to
devise a method to evaluate the traces $\Tr[ \rho U_{\theta, \phi,
\psi}]$ from experimental data. To do this, it is useful to break the integral over $\SU (1,1)$  into the sum of the contributions coming from the regions $\Omega_+, \Omega_-$ and $-\Omega_-$, introduced in  Par. (\ref{ExpoMap}). It is not difficult to see that the regions $\Omega_-$ and $-\Omega_-$ give the same contribution to the tomographic integrals, whence we have
\begin{eqnarray}\label{FeasibleInt}
\Tr[\rho A]&=& \int_{\Omega_+} \d \nu (\theta, \phi,\psi)~g(\theta,\phi,\psi) f_A(\theta,\phi,\psi) \Tr[U_{\theta, \phi,\psi} \rho]\\
\nonumber && +2  \int_{\Omega_-} \d \nu (\theta, \phi,\psi)~g(\theta,\phi,\psi) f_A(\theta,\phi,\psi) \Tr[U_{\theta, \phi,\psi} \rho]~.
\end{eqnarray}  
By definition, any element in $\Omega_+$ ($\Omega_-$) can be obtained by the exponential map as $e^{i \chi \vec n \cdot \vec K}$ for some $\chi$ and $\vec n$ with $\vec n \cdot \vec n =+1$ ($-1$).
In addiction, it is possible to show that any exponential $e^{i \chi \vec n \cdot \vec K}$ can be written as
\begin{equation}
e^{i \chi \vec n \cdot \vec K} =
\left \{ 
\begin{array}{ll}
V(\vec n)^\dag ~ e^{i \chi K_z}~V(\vec n)~, \qquad & \vec n \cdot \vec n = +1\\
W(\vec n)^\dag ~ e^{i \chi K_x}~W(\vec n)~, \qquad & \vec n \cdot \vec n = -1
\end{array}
\right.
\end{equation}
where $V(\vec n)$ and $W(\vec n)$ are suitable unitaries in the group
representation. A detailed proof of this result is given in the
Appendix.  Thanks to this observation, the trace $\Tr[\rho e^{i \chi \vec n\cdot \vec K} ]$ can be evaluated by performing a unitary transformation on
the state $\rho$ (either $V(\vec n)$ or $W(\vec n)$), and subsequently
by measuring one of the observables $K_z$ and $K_x$.

Finally we observe that, since a real experiment produces only a
finite array of data, the integral \eqref{FeasibleInt} has to be be
approximated by a statistical average over the experimental results
obtained by measuring a large number $N$ of identically prepared
systems. This introduces the need of a randomization in the
experimental setup, that produces the unitaries $V(\vec n), W(\vec n)$
according to some probability distribution. Notice that the most
natural choice, that would be to take $\d \rho (\vec n)$ as the
measure over the space of directions $\vec n$ induced by the invariant
measure $\d \nu (\theta,\phi,\psi)$ is not possible, since such a
measure cannot be normalized (the space of directions is noncompact).
The form of Eq.~\eqref{FeasibleInt} suggests then to take as a measure
$\d \nu (\theta,\phi,\psi)g(\theta,\phi,\psi)$, and in the example we
considered this actually works. However, it may happen that
regularizing the integral in Eq.~\eqref{tomid} is not sufficient for
regularizing also the group measure. In this case it is convenient to
modify $g(\theta,\phi,\psi)$ in such a way that both the measure
itself and the group integrals converge. This implies in particular
that the choice $g(\theta,\phi,\psi)\equiv1$ for square-summable
representations has to be changed. Finally, the ensemble average
$\<A\>$ can be then be approximated by the expression
\begin{equation}\label{statist}
  \<A\>\simeq\frac1N\sum_{j=1}^Nf_A(\theta_j,\phi_j,\psi_j)\Tr[\rho U(\theta_j,\phi_j,\psi_j)],
\end{equation}
where $\theta_j,\phi_j,\psi_j$ are the randomly extracted parameters.
Notice that the expression on r.h.s. in Eq.~\eqref{statist} reasonably
converges to l.h.s. if the variance of the processing function is
finite, namely if $f_A(\theta,\phi,\psi)$ is square summable. By
Eqs.~\eqref{frame} and \eqref{procefun} this condition is equivalent
to
\begin{equation}
\int_{\SU(1,1)}\d\nu(\theta,\phi,\psi)g(\theta,\phi,\psi)|f_A(\theta,\phi,\psi)|^2=\bb A|F^{-1}|A\kk<\infty.
\end{equation}

\section{conclusions}\label{conc}

This paper collects a large number of useful results about the group
$\SU(1,1)$ that are dispersed in the literature, and also contains
some novel applications regarding the use of $\SU (1,1)$ for quantum
computation and tomography with nonlinear optics. The main issues we
addressed here are \emph{i)} the approximation of $\SU(1,1)$ gates in
the quantum optical representations and \emph{ii)} the tomographic
state reconstruction exploiting group theoretical methods. As regards
the first topic, we gave an approximability theorem and discussed the
limits under which it holds.  The theorem provides a useful result in
the search for an elementary set of gates that can be used to
universally approximate any $\SU(1,1)$ gate with arbitrary accuracy.
To complete the analogy with the Solovay-Kitaev theorem for qubit
gates, the power law of the number of elementary gates as a function
of the accuracy should be evaluated, and due to non compactness we
expect that the law would depend on the parameters of the target group
element.\par

In the context of quantum estimation and tomography, we showed a
technique for regularization of the group integral for the physically
relevant representations, that are not square-summable. The core of
the regularization technique is a modification of the Haar measure over
the group, such that the regularized measure is no longer invariant.
This makes the integrals for tomographic reconstruction convergent,
but radically modifies the processing functions. Such a regularization technique is very powerful,
since it contains a freedom in the choice of the regularization
factor, that allows for a further optimization of the processing.
Moreover, the mentioned scheme can be applied not only to the case of
$\SU (1,1)$, bus also to any other tomographic setup.

\acknowledgments This work has been supported by Ministero Italiano
dell'Universit\`a e della Ricerca (MIUR) through FIRB (bando 2001) and
PRIN 2005.

\section{Appendix}
For a given representation of the $su(1,1)$ algebra, consider the real vector space $\spc V$ spanned by the generators $K_x, K_y, K_z$. Of course, $\spc V$ is isomorphic to $\Reals^3$ via the correspondence
\begin{equation}
K_x \leftrightarrow \begin{pmatrix} 1\\ 0\\ 0 \end{pmatrix} \quad K_y \leftrightarrow \begin{pmatrix}0\\1 \\0
\end{pmatrix} \quad K_z \leftrightarrow \begin{pmatrix}0\\0\\1\end{pmatrix}~.
\end{equation}  
The action of the group $\SU {(1,1)}$ on the space $\spc V$, given by $ \spc V \ni m \mapsto e^{i \chi \vec n\cdot \vec K} ~m~ e^{-i \chi \vec n\cdot \vec K}$, can be obtained by exponentiating  the adjoint action on the algebra, namely
\begin{equation}
e^{i \chi \vec n\cdot \vec K} ~m~ e^{-i \chi \vec n\cdot \vec K} = e^{i \chi \vec n \cdot {\rm Ad} (\vec K)}~m~,
\end{equation}
where ${\rm Ad}(K_i)$ is defined by ${\rm Ad} (K_i)K_j\doteq[K_i,K_j]$. Moreover, using the commutation relations of $su (1,1)$ it is immediate to find that 
\begin{equation}\label{MatrixRep}
{\rm Ad} (K_x) = 
\begin{pmatrix}
0&0&0\\
0&0&-i\\
0&-i &0
\end{pmatrix}
\quad {\rm Ad} (K_y) =
\begin{pmatrix}
0&0&i\\
0&0&0\\
i&0&0
\end{pmatrix}
\quad 
{\rm Ad} (K_z)=
\begin{pmatrix}
0&-i&0\\
i&0&0\\
0&0&0
\end{pmatrix}~.
\end{equation}  
Therefore, we obtain that a generic element of $\SU {(1,1)}$---parametrized as  $M(\theta,\phi,\psi)= e^{i(\phi-\psi)k_z}~e^{-2i k_x}~ e^{i(\phi +\psi)k_z}$ as in Eq. (\ref{Decomposition})---is represented in the space $\spc V$ by the matrix
\begin{equation}\label{SO^+(2,1)rep}
R(\theta,\phi,\psi) =  e^{i(\phi-\psi){\rm Ad} (K_z)}~e^{-2i {\rm Ad} (K_x)}~ e^{i(\phi +\psi) {\rm Ad} (K_z)} 
\end{equation} 
whose explicit expression is rather lengthy, but easily computable by
exponentiating the matrices in Eq. (\ref{MatrixRep}).

It is not difficult to see that the matrices $R (\theta, \phi, \psi)$
given by Eq. (\ref{SO^+(2,1)rep}) form a subgroup of the group $\SO
{(2,1)}$, namely they all have unit determinant and preserve the form
$\vec v \cdot \vec w = v_z w_z -v_x w_x-v_y w_y$. More precisely, the
matrices $R (\theta, \phi, \psi)$ coincide with the group $\SO^+
{(2,1)}$, which contains all matrices $R \in \SO {(2,1)}$ such that
$R_{33} \ge 1$.  Incidentally, we notice that the correspondence $\SU
{(1,1)} \to \SO^+ {(2,1)}$ is not one-to-one, since both $\pm \openone
\in \SU {(1,1)}$ are mapped into the identity in $\SO^+{(2,1)}$. One
has indeed the group homeomorphism $\SO^+{(2,1)} \simeq \SU{(1,1)}/
\mathbb{Z}_2$\cite{pererr}, which is exactly the same relation
occurring between the groups $\SU (2)$ and $\SO (3)$, namely $\SO (3)
\simeq \SU(2) /\mathbb{Z}_2$.

Similarly to the case of $\SO (3)$, where any spatial direction $\vec n$ can be conjugated with the direction of the $z-$axis by a suitable rotation, in the case of $\SO^+ (2,1)$ any direction $\vec n$ with $\vec n \cdot \vec n =+1$ can be conjugated with the $z-$axis, and any direction with $\vec n \cdot \vec n =-1$ can be conjugated with the $x-$axis.
For example, the matrix $R (\theta, \phi, \psi)$ in Eq. (\ref{SO^+(2,1)rep}) transforms the direction of the $z-$axis as
\begin{equation}
\vec k=\begin{pmatrix}
0 \\0 \\1
\end{pmatrix}
\longmapsto \vec n=
\begin{pmatrix}
-\sinh (2\theta) \sin (\phi-\psi) \\
-\sinh (2 \theta) \cos (\phi-\psi) \\
\cosh(2\theta)
\end{pmatrix}~,
\end{equation}
and it is clear that here $\vec n$ can be any direction with $\vec n\cdot\vec n =1$ (modulo an overall phase factor).
 Therefore we have, for any $\vec n$ with $\vec n\cdot\vec n =+1$
\begin{equation}
\vec n \cdot \vec K = U^\dag_{\theta, \phi, \psi} ~ K_z~U_{\theta, \phi,\psi}~, \end{equation} 
for suitable $\theta, \phi, \psi$.
In conclusion, 
\begin{equation}\label{conjug}
e^{i \chi \vec n \cdot \vec K} =U^\dag_{\theta, \phi, \psi}~e^{i \chi K_z}~ U_{\theta, \phi, \psi}~. 
\end{equation}
The same reasoning holds for any direction with $\vec n \cdot \vec n=-1$.


\begin{thebibliography}{50}
\bibitem{NotSimplyConnected} For example, the unit circle given by
  $x=y=0$, $t^2 +z^2 =1$ cannot be contracted to a point. Similarly to
  $U(1)$, the group $\SU(1,1)$ is infinitely many times connected, and
  its covering group is made by sewing together infinitely many
  sheets, each of them is homeomorphic to $\SU(1,1)$.
\bibitem{Unimodular} A unimodular group $\grp G$ is a group that
  admits an invariant Haar measure $\d \mu (g)$, namely a measure
  which is both left- and right-invariant: $\d \mu (hg) = \d \mu(gh)
  =\d \mu(g) \quad \forall g,h \in \grp G.$ For an irreducible unitary
  representation of an unimodular there are only two alternatives: the
  integral \eqref{DefSquareSummable} converges either for any vector
  $|v\> \in \spc H$, or for none (see, for example \cite{grossmorl}).
\bibitem{pererr} This fact is not mentioned in \cite{pere}, where
  actually the correspondence is mistakenly stated as $\SO{(2,1)}
  \simeq \SU{(1,1)}/ \mathbb{Z}_2$.
\end{thebibliography}
\end{document}